# Direct measurement of topological invariants through temporal adiabatic evolution of bulk states in the synthetic Brillouin zone

Zhao-Xian Chen[1,*], Yuan-hong Zhang[2,*], Xiao-Chen Sun[1,*], Ruo-Yang Zhang[3,*], Jiang-Shan Tang[1], Xin Yang[2,†], Xue-Feng Zhu[4,†] and Yan-Qing Lu[1,†]

[1]*National Laboratory of Solid State Microstructures, Collaborative Innovation Center of Advanced Microstructures, and College of Engineering and Applied Sciences, Nanjing University, Nanjing 210093, China*

[2]*College of Electrical and Information Engineering, Hunan University, Changsha 410082, People's Republic of China*

[3]*Department of Physics, The Hong Kong University of Science and Technology, Hong Kong, China*

[4]*School of Physics and Innovation Institute, Huazhong University of Science and Technology, Wuhan, Hubei 430074, China*

[*] These authors contributed equally to this work.
[†]Corresponding authors: X.Y. (xyang@hnu.edu.cn); X.Z. (xfzhu@hust.edu.cn); Y.L. (yqlu@nju.edu.cn);

*Abstract*: Mathematically, topological invariants arise from the parallel transport of eigenstates on the energy bands, which, in physics, correspond to the adiabatic dynamical evolution of transient states. It determines the presence of boundary states, while lacking direct measurements. Here, we develop time-varying programmable coupling circuits between acoustic cavities to mimic the Hamiltonians in the Brillouin zone, with which excitation and adiabatic evolution of bulk states are realized in a unit cell. By extracting the Berry phases of the bulk band, topological invariants, including the Zak phase for the SSH model and the Chern number for the AAH model, are obtained convincingly. The bulk state evolution also provides insight into the topological charges of our newly developed non-Abelian models, which are also verified by observing the adiabatic eigenframe rotation. Our work not only provides a general recipe for telling various topological invariants but also sheds light on transient acoustic wave manipulations.

For topological systems, the geometric phase, or Berry phase, results from the adiabatic evolution of the quantum state in the Brillouin zone (BZ), capturing the overall properties of the wave function rather than the minutiae and details [1,2]. This phase has significantly impacted both condensed matter physics and the steering of classical waves, including acoustic and photonic systems [3-7]. For 1D systems with certain symmetries, such as the Su-Schrieffer-Heeger (SSH) model, the geometric phase quantized into the Zak phase, a bulk topological invariant that dictates the presence of boundary states via the bulk-boundary correspondence [8,9]. In 2D systems, the Berry phase accumulated in one direction can vary along the second dimension, unveiling the topological invariant, i.e., the Chern number [10,11]. Recent advancements have extended these principles into the non-Abelian (NA) domain, where a single topological charge characterizes multiple gaps, and the boundary states in the gaps are determined by the quotient of topological invariants of the adjoining systems, conforming to NA commutation relations [12-17]. In view of the significance of topological invariants, continuous investigations have been conducted in systems such as cold atoms [18,19], waveguide lattices [20-22], and synthetic dimensions [23,24]. However, direct measurement of the topological invariants through adiabatic evolution remains a tough challenge.

Notably, the bulk state's adiabatic evolution bears multiple significance for telling the topological property [8,9]. Primarily, this method of studying the geometric phase is universally applicable and consistent with the theoretical definition across various types of topological invariants, irrespective of the specific system. Furthermore, beyond merely capturing the total geometric phase, adiabatic evolution reveals intricate details that illuminate the physical origins of topological phases. However, in both quantum and classical wave systems, observing the adiabatic evolution of wave functions poses substantial difficulties. To circumvent the complications introduced by time-dependent modulations, static waveguide systems under the paraxial approximation have been extensively utilized to mimic the time-dependent Schrödinger dynamics, facilitating studies on various phenomena such as adiabatic pumping [25-27] and Floquet topology [28-31]. Nonetheless, monitoring the transient adiabatic evolution of Bloch waves in $k$-space remains a formidable task.

To address these challenges, this Letter presents a novel method for the direct measurement of topological invariants across various topological models by implementing the adiabatic evolution of transient states in the synthetic $k$-space. As schematically shown in Fig. 1, a topological model's unit cell, containing $N$ elements, can be directly realized using gain-enhanced acoustic cavities with effective couplings via feedback circuits [32-35]. Particularly, by inserting a homemade transient phase modulator (TPM) into the circuits between the outmost two cavities, we introduce conjugated phase modulations to the intercell couplings, where the time period serves as the synthetic BZ. This setup allows direct excitation and maintenance of transient Bloch waves across different energy bands during adiabatic evolution. By extracting the geometric phase of the Bloch waves after adiabatically going through the synthetic BZ, we convincingly measure the Zak phases of the SSH model and the Chern numbers of the hopping-modulated Aubry-Andre-Harper (AAH) model. Furthermore, we develop a minimal NA three-site model, whose topological invariants are directly judged through the adiabatic evolution of its eigenstates, together with our newly developed gauge-independent relative eigenframe rotation method. Our approach paves the way for direct investigation of topological systems' properties and offers new avenues for transient energy manipulation through topological means.

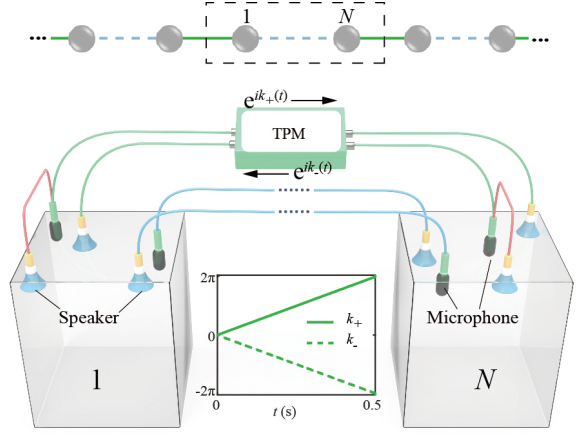

Fig. 1 Schematic for constructing a topological unit cell with time-varying couplings. The upper panel is a tight-binding model with the dashed box denoting a unit cell containing $N$ elements. The lower panel presents the acoustic implementation of this unit cell with feedback circuits in red for introducing gain to the cavities and circuits in blue (green) for realizing intracell (intercell) couplings. A dual-channel TPM is inserted in the intercell coupling circuits connecting the outmost two cavities to provide transient phase modulations $k_\pm(t) = \pm 4\pi t$ (see the inset for the measurement), serving as the synthetic BZ. Details of the experimental setup are presented in Sect. I of the Supplemental Material [36].

***Zak phases of the SSH model***. We first consider the 1D SSH model, which contains two identical elements in the unit cell and serves as the building block for many important topological models [9]. The $k$-space Hamiltonian is

$$H_{\text{SSH}}(k) = 2\pi \begin{bmatrix} f_0 & v + we^{-ik} \\ v + we^{ik} & f_0 \end{bmatrix}, \quad (1)$$

where $f_0$ is the resonant frequency of the cavities, $v$ and $w$ are the amplitudes of the intracell and intercell couplings, respectively. Utilizing the eigenvectors $|E_\pm(k)\rangle$ of the energy bands $\varepsilon_\pm(k)$, the SSH model's Zak phase can be calculated by integrating the Berry connection across the first BZ: $\beta_\pm = i \int_{BZ} \langle E_\pm(k)|\partial_k|E_\pm(k)\rangle dk$, yielding $\beta = 0$ for $v > w$ (trivial phase) and $\beta = \pi \pmod{2\pi}$ $v < w$ (nontrivial phase). Notably, upon transforming $k$ into a function of time $t$, the Berry phase manifests as the geometric phase acquired by a transient state adiabatically evolving through a whole time period. Thus, the Berry phase can be obtained by subtracting the dynamic phase from the total accumulated phase, making the adiabatic evolution a possible solution for its measurement (see Sect. II of [36]).

Experimentally, we construct the SSH model's unit cell with two acoustic cavities resonating at $f_0 = 1600$Hz. By utilizing in-phase feedback circuits, we reduce the cavities' damping rate to $\gamma = 0.8$ Hz, making the eigenstate's evolution observable in time. The couplings are defined as $v + w = 12$ Hz and $w - v = 2\Delta$, facilitating a phase transition by modifying $\Delta$. The phase modulation of the TPM in the coupling circuit is configured as $k_\pm(t) = \pm 4\pi t$ from 0 to 0.5s. To ensure the adiabaticity of the state evolution, we select $\Delta = \pm 3$Hz, obtaining substantial bandgaps across the entire BZ, as depicted in Fig. 2(a). In the measurement, we use two channel signals of $\sin(2\pi\varepsilon_\pm t)|E_\pm\rangle$ to stimulate the lower or upper eigenmode at $k = 0$ and then start the phase modulation with the sound sources switched off. The recorded sound pressures within the two cavities, presented in Fig. 2(b), facilitate the extraction of the Berry phase $\beta_\pm$. For $\Delta = -3$Hz, Fig. 2(c) shows the phase lag between the two cavities, denoted as $\arg(p_2/p_1)$, for the two eigenmodes during the phase modulation. The experimental results are in line with the stroboscopic predictions, affirming the adiabaticity of the process. The Zak phases are extracted as $\beta_\pm \approx 0$, aligning with the theoretical predictions. Similarly, we record the adiabatic evolutions of the bulk bands for the case of $\Delta = 3$Hz, with the phase lags between the two cavities shown in Fig. 2(d). In contrast to the previous case, the Zak phases are determined as $\beta_\pm \approx \pi$, underscoring the non-trivial topological nature. Notably, the model's topological essence can also be read from the bulk mode's parity at high symmetry positions in time, i.e., 0 and 0.25s in our experiment. Specifically, $\arg(p_2/p_1)$ of the two bands remain unchanged at these points for the trivial case (Fig. 2(c)), yet swap for the nontrivial case (Fig. 2(d)). In addition, the adiabaticity of the evolution process can be assessed with [37]

$$\mu_{+-}(t) = |\frac{\langle E_+(t)|\partial_t|E_-(t)\rangle}{\varepsilon_+ - \varepsilon_-}|, \quad (2)$$

which provides a general method to quantify the intermodal transition. In our system, $\mu_{+-}$ gets larger when $|\Delta|$ decreases, indicating a deterioration in adiabaticity due to the reduction of bandgap. Theoretically, the diabatic transition can be prohibited by applying a shortcut, which facilitates the acquisition of a complete topological phase diagram (see Sect. III of [36]) [38].

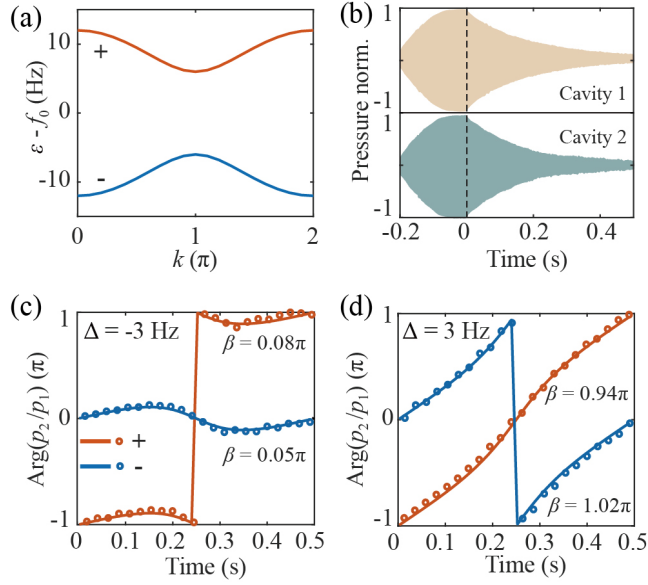

Fig. 2 (a) Energy band for the SSH model with $\Delta = \pm 3$Hz. (b) Recorded excitation ($k_\pm = 0$ for $t < 0$) and modulation ($k_\pm = \pm 4\pi t$ for $0 < t < 0.5s$) of the lower band. (c-d) Measured (marks) and theoretical (curves) phase lags between the two cavities during the modulation for $\Delta = -3$Hz (c) and $\Delta = 3$Hz (d). Extracted Zak phases are labeled accordingly.

***Chern numbers of the AAH model***. For the AAH model, in addition to the Floquet boundary condition $k$, it incorporates a second parameter $\varphi$, which modulates the on-site frequencies or off-diagonal couplings [39,40]. Thus, the AAH model possesses a 2D parameter space, and the topology of a bulk band can be characterized with the Chern number, obtained through the 2D integration of the Berry curvature:

$$C_n = \frac{i}{2\pi}\int_0^{2\pi} dk \int_0^{2\pi} d\varphi \, \nabla \times \langle E_n|\nabla_{k,\varphi}|E_n\rangle. \quad (3)$$

Additionally, we can discretize $\varphi$ and get the Berry phase $\beta_n$ along the $k$ direction, effectively reducing the AAH to one dimension. The winding of $\beta_n(\varphi)$ correlates with the Chern number [11]. To demonstrate this, we construct a hopping-modulated AAH model with three identical elements per unit cell. The Hamiltonian is expressed as

$$H_{\text{AAH}}(k,\varphi) = 2\pi \begin{bmatrix} f_0 & v_1(\varphi) & v_3(\varphi)e^{-ik} \\ v_1(\varphi) & f_0 & v_2(\varphi) \\ v_3(\varphi)e^{ik} & v_2(\varphi) & f_0 \end{bmatrix}, \qquad (4)$$

where $v_j(\varphi) = v_0 + v_m \cos(2\pi j/3 + \varphi)$ for $j = 1,2,3$, representing the couplings between adjacent sites. This unit cell is physically realized with three acoustic cavities with the couplings schematically shown in Fig. 3(a). We set $v_0 = v_m = 5$ Hz, with which the energy bands of the Hamiltonian are calculated and shown in Fig. 3(b) as a function of $k$ and $\varphi$. Using Eq. (3), the Chern numbers of the bulk bands are calculated as $1$, $-2$ and $1$, respectively. Notably, at each $\varphi$ point, the average bulk band energies satisfy $\bar{\varepsilon}_n = \varepsilon_n(k = -0.5\pi)$, implying the same dynamic phases when the bulk state $|E_n(k = -0.5\pi)\rangle$ traverses the entire BZ or remains unmodulated for equivalent durations. Consequently, we can easily extract the Berry phase along the $k$ direction by using $|E_n(k = -0.5\pi)\rangle$ as the initial state and assessing the final phase difference between the processes with and without transient phase modulation. Considering the substantial bandgaps, the adiabatic evolution of states is ensured, as the condition $\mu \ll 1$ is satisfied for all interband transitions (see Fig. 3(c) for $\varphi = 0$). This is further corroborated by comparing the transient phase lag variation with stroboscopic prediction (see Fig. S5 of [36]). We discretize $\varphi$ into 20 points and extract $\beta_n$ as a function of $\varphi$ for the three bands. As shown in Fig. 3(d) and 3(f), $\beta(\varphi)$ for bands I and III exhibit one full winding with $d\beta/d\varphi \geq 0$, consistent with the Chern number $C_{1,3} = 1$. In contrast, $\beta(\varphi)$ in Fig. 3(e) winds twice with $d\beta/d\varphi \leq 0$, echoing with the Chern number $C_2 = -2$. Remarkably, all the measurements are in line with theoretical predictions. Notably, we have $\beta_n = 0$ for all the three bands at $\varphi = \pi$, which corresponds to no modulation along the $k$ direction and results in flat energy bands (see Fig. S5 of [36]). Our method for extracting Berry phases is universally applicable and can be extended to other 2D topological models, such as Haldane model and Kane-Mele model [41,42].

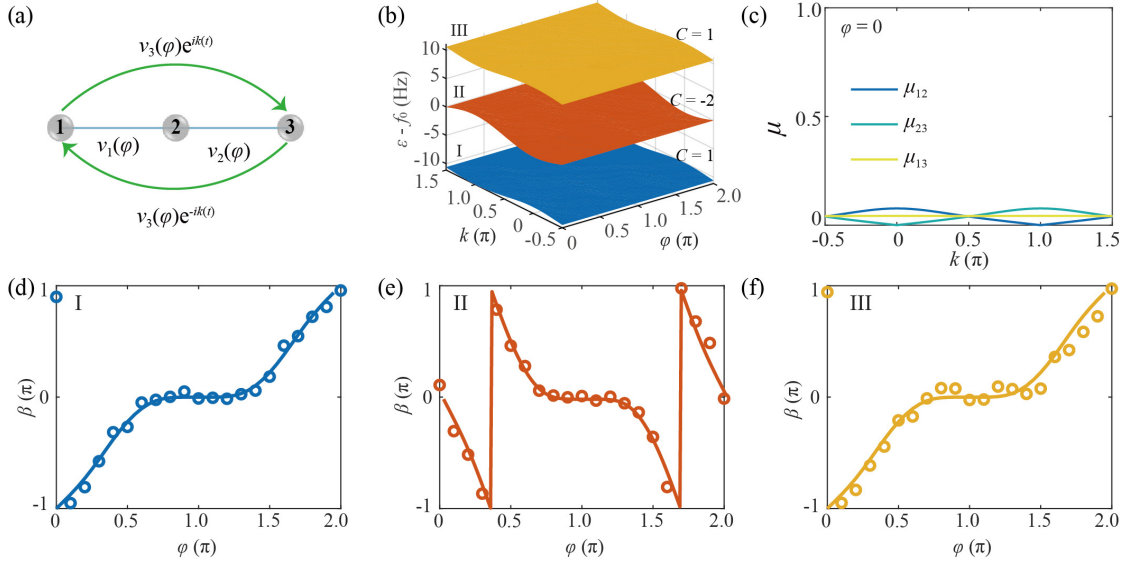

Fig. 3 (a) Tight-binding configuration for the unit cell within the hopping-modulated AAH model. (b) Energy bands of the unit cell as a function of $k$ and $\varphi$, with the bulk Chern numbers annotated accordingly. (c) Diabatic transition between every two bands with $\varphi = 0$. Phase modulations are designed as $k_{\pm}(t) = -0.5\pi \pm 4\pi t$. (d-f) Extracted (open circles) and calculated (solid curves) Berry phases for the three bands as functions of $\varphi$, reflecting the Chern numbers of $-2$ for the second band (e) and $+1$ for the other two bands (d and f), respectively.

*Topological charges for a non-Abelian model*. So far, we have been focusing on Abelian topological models with every bulk band considered separately. When extending to parity-time (PT) symmetric systems with multiple bandgaps, the topological phases are characterized by NA topological charges [7,12,13]. These charges can be determined with the eigenstates' rotation on the eigenstate sphere. However, previous approaches have relied on the real gauge of PT-symmetric Hamiltonians under specific bases. In what follows, we will introduce a gauge-independent method that identifies NA topological charges through the adiabatic evolution path in the $SO(3)$ manifold.

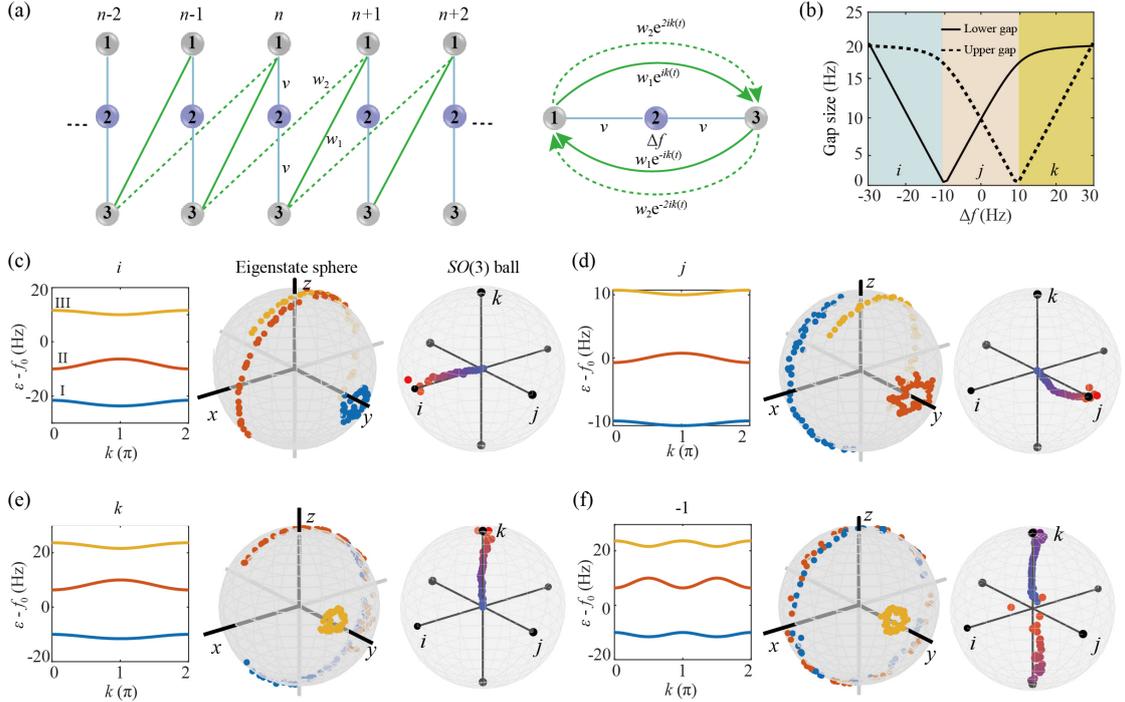

Fig. 4 (a) The left panel illustrates the tight-binding model of an NA three-band system, while the right panel depicts the implementation within a unit cell. (b) NA phase diagram of the system as a function of $\Delta f$. The solid and dashed curves represent the minimal widths of the two gaps. Other parameters are given in the main text. (c) The left panel displays the energy bands of the system carrying a quaternion charge of $i$. The middle panel showcases the eigenstate trajectories on a sphere, where the first eigenstate (in blue) completes a full cycle upon adiabatically traversing the BZ. Conversely, the second and third eigenstates (in brown and yellow) rotate to the antipodal points of the initial ones. The topological charge can also manifest as the adiabatic trajectory $R(k)$ in $SO(3)$ ball, with the color from blue to red tracing the modulation from $k=0$ to $2\pi$ (right panel). (d-f) Similar to (c) with respect to the NA topological charge $j$ (d), $k$ (e), and $-1$ (f), respectively.

For demonstration, here we come up with a PT-symmetric three-band model, as schematically shown in Fig. 4(a), which is determined by the conjugacy classes $i, j, k$ and $\pm 1$ of the NA quaternion group $\mathbb{Q}$. The design scheme is presented in Sect. V of [36], and the $k$-space Hamiltonian is

$$H_{\text{NA}}(k) = 2\pi \begin{bmatrix} f_0 & v & w_1 e^{-ik} + w_2 e^{-2ik} \\ v & f_0 + \Delta f & v \\ w_1 e^{ik} + w_2 e^{2ik} & v & f_0 \end{bmatrix}, \quad (5)$$

where $\Delta f$ is the offset of the second cavity, $w_1$ and $w_2$ are the nearest neighbor and next-nearest neighbor couplings, respectively. In the following analysis, we keep the intracell coupling $v$ smaller than $w_1$ or $w_2$. This configuration allows us to conceptualize the NA model as an insertion of an offset cavity into a nontrivial SSH model. We first ignore $w_2$ and set $v = 2\text{Hz}$ and $w_1 = 10\text{Hz}$ to investigate the phase transition as a function of $\Delta f$. As depicted in Fig. 4(b), by changing $\Delta f$ from $-30$ to $30\text{Hz}$, we observe the emergence of three distinct NA phases $i, j$ and $k$. These phase transitions are marked by the closing and reopening of energy gaps. In the experiments, we construct three acoustic cavities with dynamic couplings to simulate the NA Hamiltonian, as illustrated in the right panel of Fig. 4(a). We first select three different values of $\Delta f$ to realize the NA model in the $i$, $j$, $k$ phases, respectively (see Figs. 4(c-e)). Additionally, we achieve the "$-1$" NA phase, as depicted in Fig. 4(f), by disregarding $w_1$ and setting $w_2 = 10\text{Hz}$ (see Fig. S7 in [36]). Akin to the Abelian scenarios, we let the transient states initially reside at the left end ($k = 0$) of each band and examine the adiabatic evolution of the transient states on all bands, from which the NA topological charges can be identified by two different means.

The first method leverages PT symmetry to convert the Hamiltonian into a real matrix via a unitary transformation, $H_r(k) = U^\dagger H_{NA}(k) U$. Accordingly, the frame of eigenstates $\Phi(k) = (|E_1(k)\rangle, |E_2(k)\rangle, |E_3(k)\rangle)$ transforms into a real orthogonal matrix $\Phi_r(k) = U^\dagger \Phi(k)$ and hence can be mapped to three orthogonal vectors on the unit sphere. By tracing the transformed real eigenvectors $\Phi_r(k)$ on the sphere throughout a complete adiabatic evolution period, we discern unique paths for different NA phases, as demonstrated in the middle panels of Figs. 4(c-f). For the cases of topological charges $q = i, j, k$, there is always one closed trajectory, while the other two form open paths each connecting a pair of antipodal points. Depending on whether the closed trajectory is the first, second, or third band determines whether the topological charge is $i$, $j$, or $k$. In contrast, for $q = -1$, all three trajectories form closed loops on the sphere with an accumulated rotation angle of $2\pi$. All these experimental observations are consistent with the theoretical predictions.

While the first method requires artificially transforming the transient states into the real gauge at every moment, we have also developed a new approach to efficiently identify the NA topological phases in an arbitrary complex gauge, with the detailed derivations presented in Sect. V of [36]. During an adiabatic evolution, we may consider the "relative rotation" of the transient states $\Phi(k)$ with respect to the initial one: $R(k) = \Phi(0)^{-1}\Phi(k)$, where the dynamic phases have been removed from $\Phi(k)$. It can be shown that provided the phases of initial states are selected properly, $R(k)$ is guaranteed to be a real orthogonal matrix belonging to $SO(3)$ group throughout the adiabatic evolution, irrespective of the gauge of the Hamiltonian. Moreover, for the cases of $q = i, j, k$, the final states of $R(k)$ converge to three fixed rotations, i.e., $R_i(2\pi) = \hat{\mathcal{R}}(\pi, \hat{\mathbf{x}})$, $R_j(2\pi) = \hat{\mathcal{R}}(\pi, \hat{\mathbf{y}})$ and $R_k(2\pi) = \hat{\mathcal{R}}(\pi, \hat{\mathbf{z}})$, where $\hat{\mathcal{R}}(\theta, \hat{\mathbf{n}})$ represents a $SO(3)$ rotation with $\theta$ and $\hat{\mathbf{n}}$ being the rotation angle and the rotation axis, respectively. In the right panels of Figs. 4(c-f), a ball with radius $r = \pi$ is used to represent the $SO(3)$ manifold. Every point $\mathbf{p}$ inside the ball indicates a $SO(3)$ rotation $\hat{\mathcal{R}}(|\mathbf{p}|, \hat{\mathbf{p}})$. And two antipodal points on the sphere give the same rotation $\hat{\mathcal{R}}(\pi, \hat{\mathbf{p}}) = \hat{\mathcal{R}}(-\pi, -\hat{\mathbf{p}})$. Consequently, the NA charges $q = i, j, k$ manifest as three inequivalent adiabatic open paths $R_{i,j,k}(k)$ in the $SO(3)$ ball, each connecting the origin (identity matrix) and an endpoint of one of the three axis $i, j, k$, respectively. Meanwhile, the NA charge $q = -1$, characterized by an accumulated rotation angle of $2\pi$, presents as a closed path starting at the origin, traversing the entire $SO(3)$ ball through a pair of antipodal points on the sphere and returning to the origin at the end of the adiabatic cycle. We emphasize that this gauge-independent manifestation of NA quaternion charges is generally applicable to any PT-symmetric three-band systems.

To sum up, our research has successfully formulated an accessible approach for observing the transient adiabatic bulk state evolution in a time-varying acoustic unit cell controlled with feedback circuits, providing a fertile testbed for time-modulation related research, such as Landau-Zener tunneling [43] and non-reciprocal effects [44]. The topological invariants for Abelian systems, including the SSH model and the AAH model, have been directly observed by extracting the Berry phases from the transient state evolutions along the bulk bands, which can be further extended to non-Hermitian or Floquet topological systems [45,46]. Beyond the Abelian scenarios, our work stands out by introducing a gauge-independent approach for identifying NA topological charges, validated through detecting the adiabatic evolution of a PT-symmetric three-band model. This complements the traditional methods that only work for the real gauge, offering deeper insights into NA topological systems and potentially applicable to other NA braiding processes [47-49]. The insights gained from our study may also contribute to the design of novel devices with robust functionalities, such as topological transistors and logic gates [50,51].

*Acknowledge*. This work was supported by the National Key R&D Program of China (No. 2022YFA1405000), the Natural Science Foundation of Jiangsu Province (No. BK20212004), the National Natural Science Foundation of China (No. 52103341), the China Postdoctoral Science Foundation (No. 2023M731609), Jiangsu Funding Program for Excellent Postdoctoral Talent (No. 2023ZB473), the Croucher Foundation (No. CAS20SC01), and a grant from the Department of Science and Technology of Hunan Province, China (No. 2019RS1028).